\documentstyle[12pt]{article}
\begin{document}
\vsize=25.0 true cm
\hsize=16.0 true cm
\predisplaypenalty=0
\abovedisplayskip=3mm plus 6pt minus 4pt
\belowdisplayskip=3mm plus 6pt minus 4pt
\abovedisplayshortskip=0mm plus 6pt
\belowdisplayshortskip=2mm plus 6pt minus 4pt
\normalbaselineskip=14pt
\normalbaselines
\begin{titlepage}
 
\begin{flushright}
{\bf CERN-TH/99-121}
\end{flushright}
 
\vspace{0.5cm}
\begin{center}
{\bf\Large Library of
anomalous $\tau\tau\gamma$ couplings for 
$\tau^+\tau^- (n\gamma)$ \\ Monte Carlo programs
}\end{center}
 \vspace{0.3cm}
\begin{center}
  {\bf T. Paul} and {\bf J. Swain} \\
  {\em Department of Physics, Northeastern University, Boston, MA 02115, USA}\\

and\\
\vspace{0.2cm}
   {\bf Z. W\c{a}s } \\
  {\em CERN, Theory Division, CH 1211, Geneva 23, Switzerland,\\ and}\\
   {\em Institute of Nuclear Physics,
        Krak\'ow, ul. Kawiory 26a, Poland}\\
\end{center}
 
\vspace{2.5cm}
\begin{center}
{\bf   ABSTRACT}
\end{center}
We briefly describe a library that may be used with any $e^+e^- \to \tau^+ \tau^- (n\gamma)$
Monte Carlo program to account for the effects of anomalous $\tau \tau \gamma$ couplings.
The implementation of this library in KORALZ version 4.04 is discussed.  

\vskip 2.9 cm
\centerline{\it To be submitted to Computer Physics Communications}
\vspace{0.4cm}

\begin{flushleft}
{\bf 
 CERN-TH/99-121\\
May 1999}
\end{flushleft}
\footnoterule
\noindent
{\footnotesize
\begin{itemize}
\item[${\dagger}$]
Work supported in part by 
the US National Science Foundation and 
Polish Government grants 
KBN 2P03B08414, 
KBN 2P03B14715, 
Maria Sk\l{}odowska-Curie Joint Fund II PAA/DOE-97-316,
and Polish-French Collaboration within IN2P3.
\end{itemize}
}
 
\end{titlepage}

 \vskip 10pt
 
 
\section{Introduction}

Radiative $\tau$ pair production is of great interest, as it is 
sensitive to anomalous electromagnetic couplings of the
$\tau$.  With the sensitivity afforded by the LEP experiments,
this provides an opportunity to search for new physics 
phenomena~[1--4].
Any meaningful interpretation of the experimental data requires 
a Monte Carlo simulation in which Standard Model predictions
may be augmented by the contributions from possible anomalous couplings.

Since the LEP collaborations are entering their final years of operation
it is a good time to document the programs that were actually used in 
data analyses.  In this paper we describe a library that has been used to
calculate anomalous contributions to $\tau\tau\gamma$ couplings~\cite{ttg_l3}.
The library is based on the work described in~\cite{ttg_theory} and
can be used with any 
$e^+e^- \to \tau^+ \tau^- ( n\gamma)$ Monte Carlo program and, after minor
adaptation, with $p p \to Z/\gamma + ...$; $Z/\gamma \to  \tau^+ \tau^- ( n\gamma)$
or  $e p \to Z/\gamma + ...$; $Z/\gamma \to \tau^+ \tau^- ( n\gamma)$
programs as well. 

In the present paper, we will discuss the interface of our library to
KORALZ version 4.04, which is described in detail 
in~\cite{KORALZ,KORALZ1}. 
The fortran code of the library 
is archived together with KORALZ~\cite{KORALZ1}, in the same 
tree of directories.
Let us note that in the future, KORALZ 
will be replaced by a new program, KK2f~\cite{KK2f}, which is based on 
a more powerful exponentiation at the spin amplitude level; implementation
of our library will be straightforward for that program as well.

\section{Calculation of anomalous couplings}

To evaluate the effects of anomalous electromagnetic couplings on radiative
$\tau$ pair production, a tree--level calculation of the squared matrix element 
for the process
$e^+ e^- \rightarrow \tau^+ \tau^- \gamma$ has been carried out~\cite{ttg_theory},
including contributions from the anomalous magnetic dipole moment at $q^2=0$, $F_2(0)$, 
and the electric dipole moment $F_3(0)$.           
This calculation is included in our library. When activated, it
uses the 4-momenta of the leptons and the photon generated by the host program 
to compute a weight, $w$, for each event according to
\begin{equation}
w = \frac{| {\cal M}_{\mathrm ano} | ^2 } {| {\cal M}_{\mathrm SM} | ^2}.
\end{equation}
${\cal M}_{\mathrm ano}$ is the matrix element for $F_2(0) \neq 0$ and/or 
$F_3(0) \neq 0$, and ${\cal M}_{\mathrm SM}$ is the matrix element for 
$F_2(0) = F_3(0) = 0$.

As this calculation is performed at ${\cal O} (\alpha)$, 
the case of multiple bremsstrahlung requires special treatment.  
In this case, a reduction procedure 
is first applied in which all photons, except the one with the greatest momentum
transverse to a lepton, or $p_T$,
are incorporated into the 4-momenta of effective initial--\, or final--state leptons.
The 4-momenta of the photon with greatest $p_T$ and the effective
leptons are then used to compute the weight.
Cross--checks of the calculation against an independent and slightly simplified
analytical calculation~\cite{Riemann} as well as checks of the validity of 
the reduction procedure are described in~\cite{Paul}.  The results of the 
calculation have been used in the measurement of anomalous electromagnetic
moments of the $\tau$ described in~\cite{ttg_l3}.

\section{Flags to control anomalous couplings in KORALZ}
In KORALZ version 4.04, the calculation in our library 
is activated by setting the card {\tt IFKALIN=2}.
This is transmitted from the main program via the KORALZ input parameter {\tt NPAR(15)}.
Additional input parameters are set in the routine {\tt kzphynew(XPAR,NPAR)}, although 
there are currently no connections to the KORALZ matrix input parameters 
{\tt XPAR} and {\tt NPAR}.  Table~\ref{tab:parameters} summarizes the functions  
of these input parameters.

\begin{table}[hbt!]
\renewcommand{\arraystretch}{2.0}

\begin{center}
{

\begin{tabular}{|l| l| c | } \hline
\setlength{\tabcolsep}{3mm}

Parameter             &      Description     & Default    \\ \hline
IFL1                  &    Compute weights for $F_2(0)$ if $\mathrm{IFL1}=1$  & 1        \\ \hline
IFL2                  &    Compute weights for $F_3(0)$ if $\mathrm{IFL2}=1$  & 0        \\ \hline
ISFL                  &     \parbox{0.7\textwidth} {\rule{0mm}{5mm}For $\mathrm{ISFL}=-1$, compute {\em only} terms with 
                                 anomalous contributions.  For $\mathrm{ISFL=0}$, include 
                                 all terms (anomalous, Standard Model, all interference).
                                 For $\mathrm{ISFL}=1$, use the approximation of ref.~\cite{Riemann}}\rule{0mm}{10mm} & 0  \\ \hline
IRECSOFT              & Generate {\em only} events with photon(s) if $\mathrm{IRECSOFT=1}$     & 0        \\ \hline
EMINACT               &    \parbox{0.7\textwidth} {Minimum sum of all photon energies required 
                                   to calculate anomalous weights}   &  17 GeV\\ \hline
EMAXACT               &    \parbox{0.7\textwidth} {Maximum sum of all photon energies allowed
                                   to calculate anomalous weights}   & 1000 GeV \\ \hline
PTACT                 &    \parbox{0.7\textwidth} {Minimum sum of all photon momenta transverse to the 
                                   beam direction required to calculate anomalous weights} & 2 GeV\\ \hline

\end{tabular}
}

\caption{Input parameters to control the calculation of weights for anomalous 
electromagnetic moments.}

\label{tab:parameters}
\end{center}
\end{table}

In order to provide the user with enough information to retrieve $w$ 
for a given event for any $F_2(0)$ or $F_3(0)$, we take advantage of 
the fact that, for each event, one may write $w$ as a quadratic function of the 
anomalous couplings:
\begin{equation} \label{eqn:weight}
w = \alpha F_2^2(0) + \beta F_2(0) + \gamma F_3^2(0) + \delta F_3(0) + \epsilon~.
\end{equation}
When the calculation of $w$ is completed, 
the 5 weight parameters $\alpha, \beta, \gamma, \delta$ and $\epsilon$ 
are stored in the common block {\tt common /kalinout/ wtkal(6)}, 
with the assignments shown in Table~\ref{tab:weights}.
\begin{table}[hbt!]
\begin{center}
\begin{tabular}{|l|l|} \hline

Common block entry   &  Weight parameter \\ \hline 

{\tt wtkal(1)}  &  not used here \\ 
{\tt wtkal(2)}  &  $\epsilon$\\
{\tt wtakl(3)}  &  $\alpha$  \\    
{\tt wtkal(4)}  &  $\beta$   \\
{\tt wtkal(5)}  &  $\gamma$  \\
{\tt wtkal(6)}  &  $\delta$  \\  \hline

\end{tabular}
\caption{Correspondence between entries in {\tt kalinout} common block entries
and weight parameters of eq.~(\ref{eqn:weight}).}

\label{tab:weights}
\end{center}
\end{table}
The user is then free to calculate $w$ for whatever combination of $F_2(0)$ 
and $F_3(0)$ is desired.
In practice we set $\epsilon=1$, since anomalous terms must vanish
for $F_2(0)=F_3(0)=0$. We also set $\delta=0$, as the interference between 
Standard Model and anomalous amplitudes vanishes in the case of 
radiation from an electric dipole moment. These short cuts save 
substantial CPU time.

The code for calculation of the weight $w$ is placed
in the directory {\tt korz{\_}new/ttglib} in the files {\tt ttg.f}
and {\tt ttgface.f}.

\section{Demonstration programs}
The demonstration program {\tt DEMO2.f} for the run of KORALZ
when our library is activated 
can be found in the directory 
{ \tt korz{\_}new/february} and the output {\tt DEMO2.out}
in the directory { \tt korz{\_}new/february/prod1}. The Standard Model {\tt DEMO.f} for KORALZ
and its output {\tt DEMO.out} are also included in the directories
mentioned above. All these files, as well as the library itself,
are archived together with KORALZ \cite{KORALZ1}.
The maximum centre-of-mass energy allowed for the runs
with anomalous $\tau\tau\gamma$ couplings is $200$~GeV.

\section*{Acknowledgements}
ZW would like to thank the L3 group of ETH Z{\"{u}}rich for support
while this work was performed.


\end{document}